\documentclass[]{raa}            

\usepackage{graphicx,times}             
\usepackage{natbib}

\begin{document}

\title{Binary Population  Synthesis}

\volnopage{Vol.0 (200x) No.0, 000--000}      
\setcounter{page}{1}          

\author{Zhanwen Han
\inst{1,2,3,4}
\and Hongwei Ge
\inst{1,2,3,4}
\and Xuefei Chen
\inst{1,2,3,4}
\and Hailiang Chen
\inst{1,2,3,4}
   }
\institute{Yunnan Observatories, Chinese Academy of Sciences, Kunming 650216,  China; {\it zhanwenhan@ynao.ac.cn, gehw@ynao.ac.cn, cxf@ynao.ac.cn, chenhl@ynao.ac.cn}\\
\and
Key Laboratory for the Structure and Evolution of Celestial Objects, Chinese Academy of Sciences, Kunming 650216,  China;\\
\and
University of the Chinese Academy of Sciences, Beijing 100049, China.\\
\and
Center for Astronomical Mega-Science, Chinese Academy of Sciences, Beijing 100012, China\\
}
   \date{Received~~2009 month day; accepted~~2009~~month day}

\abstract{
Binary interactions lead to the formation of intriguing objects, such as compact binaries, supernovae, gamma ray bursts, X-ray binaries, pulsars, novae, cataclysmic variables, hot subdwarf stars, barium stars, and blue stragglers. To study the evolution of binary populations and the consequent formation of these objects, many methods have been developed over the years, of which a robust approach named binary population synthesis (BPS) warrants special attention. This approach has seen widespread use in many areas of astrophysics, including but not limited to analyses of the stellar content of galaxies, research on galactic chemical evolution, and studies concerning star formation and cosmic re-ionization. In this review, we discuss the role of BPS, its general picture, and  the various components that comprise it. We  pay special attention to the stability criteria for mass transfer in binaries, as this stability largely determines the fate of binary systems. We conclude with our perspectives regarding the future of this field.
}

   \authorrunning{Han et al.\ }            
   \titlerunning{Binary Population Synthesis}  

   \maketitle

%
%

\section{Introduction}

Stars are building blocks of galaxies and the Universe, 
and most of what we know about the Universe comes from stars.  
The theory of stellar structure and evolution is one of the cornerstones of astrophysics.
It is based on assumptions of hydrostatic equilibrium and what we know about the energy supply coming
from thermonuclear reactions,  
and enhances our understanding of the  internal  physical structure and evolution of stars.
To develop the theory of stellar structure and evolution, astronomers
developed many computer codes in the 60s, e.g.
\cite{Henyey1959}, \cite{Iben1965}, \cite{Kippenhahn1967}, \cite{Paczynski1970}, \cite{Eggleton1967, Eggleton1971, Eggleton1972, Eggleton1973a}, \cite{Eggleton1973b}.
Many more were invented later, e.g. the Geneva code (\citealt{Maeder1987}), the PAdova and TRieste Stellar Evolution Code (\citealt{Bressan1993}), EZ code (\citealt{Paxton2004}), 
with MESA (\citealt{Paxton2011}) being the latest and arguably the most robust. These codes solve the basic equations, i.e.
the mass continuity equation, the hydrostatic equilibrium equation, 
the energy conservation equation, 
the energy transport equation and the chemical composition equation, 
to obtain the structure and evolution of a star. The observed phenomena and the general laws that govern the stellar world are very well explained and reproduced with the help of those codes. These codes have driven the development of many aspects of astrophysics. 

In the 1990s, however, our understanding of stellar structure and evolution faced many challenges. (i) Some stars display characteristics apparently contrary to predictions made by stellar structure and evolution theory. These exotic stars are generally used to probe evolutionary processes in stars, to derive ages and metallicities of stellar populations in galaxies, and to measure cosmological distances. Such exotic stars therefore play a crucial role in our understanding of stellar physics, the structure and evolution of galaxies, and cosmology, and they have been a key subject of study for many decades. (ii) At least  half  of all stars are in  binaries  (two stars  orbiting  each other due to gravitation). Binary interaction  makes stellar evolution more  complicated, and leads to the existence of most exotic stars and strange observational phenomena. Some basic problems in binary evolution have not been resolved yet. (iii) Large surveys, such as the Sloan Digital Sky Survey (SDSS) (\citealt{York2000}), 
Large Area Multi-Object Fiber Spectroscopic Telescope (LAMOST) survey 
(\citealt{Zhao2012}), Gaia (\citealt{Gaia2016}), Kepler (\citealt{Borucki2010}) and TESS (\citealt{Ricker2014}), have revealed the statistical properties of stars and galaxies,  and are making 
significant  progress  in  astrophysics. However, traditional  stellar evolution theory can only evolve one star at a time (single or binary) and are therefore not able to explain the statistical properties of stellar populations. The importance  in comparing theory with observation is that real physical processes are then revealed. 
(iv)  A galaxy consists of thousands of millions of stars,  and the study of its evolution requires knowledge of the evolution of stellar populations. With the advent of galaxy and cosmology studies,  it becomes urgent to develop an approach to evolve millions of stars at the same time.

To resolve these issues, astronomers have developed binary population synthesis (BPS), e.g. \cite{deKool1992,deKool1993}, \cite{Yungelson1993, Yungelson1994}, \cite{Han1994,Han1995a,Han1995b}. BPS is a robust approach
to evolve a large number of stars (including binaries) so that we can explain, understand and predict the properties of a population of a type of stars. With BPS, we are able to unveil the underlying crucial physical processes and explore the scenarios 
for the formation and evolution of those exotic stars.
Many BPS codes have been developed, e.g. Scenario Machine (\citealt{Lipunov1996,Lipunov2007}), SeBa 
(\citealt{PortegiesZwart1996,Nelemans2001a,Toonen2012}),  
Yunnan Model (\citealt{Han1998,Han2002,Han2003b,Zhang2002,Zhang2004,Zhang2005}), BSE (\citealt{Hurley2002}), 
StarTrack (\citealt{Belczynski2002a}), BiSEPS (\citealt{Willems2002}), BPASS (\citealt{Eldridge2008}), 
SYCLIST (\citealt{Georgy2014}), COMPAS (\citealt{Stevenson2017}), MOBSE (\citealt{Giacobbo2018}), Combine (\citealt{Kruckow2018}), dart-board (\citealt{Andrews2018}), COSMIC (\citealt{Breivik2020}),
etc. Those codes are widely used in the study of many aspects of astrophysics.  BPS is now a common practice in the investigation of exotic objects, e.g. double black holes, double neutron stars, double white dwarfs, supernovae, gamma ray bursts, X-ray binaries, pulsars,
novae, cataclysmic variables, hot subdwarfs stars, barium stars, blue stragglers, etc., and is also used in many other areas, e.g. in the studies of spectral 
energy distribution of galaxies, and the chemical evolution of galaxies.

\section{The Role of Binary Population Synthesis}

Binary population synthesis (BPS) studies play a significant role in many aspects of astrophysics, from physical processes in stellar evolution and binary evolution, the formation of binary related objects, to evolutionary population synthesis, galaxy evolution and re-ionization of the Universe.

Indeed, the theory of stellar evolution has achieved great success, but many problems have not been resolved yet due to the complexity of the problems. One such example is the third dredge up of thermally pulsating asymptotic giant branch (TPAGB) stars and the production of $s$-elements. The dredge up and the surface abundances of $s$-elements depend very sensitively on the detailed numerical treatment of the burning shells. Carbon-enhanced metal-poor (CEMP) stars are enriched in $s$-process elements, and are formed via mass-transfer of carbon-rich material from a TPAGB primary star to a less massive main-sequence companion which is seen today. By comparing the results of BPS studies of CEMP stars to that of observations, \cite{Izzard2009} constrained the mass range for stars undergoing efficient 3rd dredge up (see also \citealt{Han1995a}). \cite{Han2002,Han2003b} carried out a detailed and systemtic BPS study on hot subdwarf stars and constrained to a great extent the common envelope (CE) evolution of binary stars. \cite{Chrimes2020} inferred the core angular momentum threshold for jet production of collapsars by studying long-duration gamma-ray bursts (GRBs) via BPS. BPS provides an approach to tackle problems which are difficult to tackle otherwise.    

Type Ia supernovae have been successfully used as a cosmological distance indicator, leading to the discovery of the accelerated expansion of the Universe and consequently the inferred existence of dark energy
(\citealt{Riess1998,Perlmutter1999}). Recent Hubble constant measurements from type Ia supernovae disagree at $4\sigma$ to $6\sigma$ with that from Planck observations of the Cosmic Microwave Background in conjunction with the standard cosmological model (\citealt{Planck2018,Riess2019}), causing a crisis in cosmology. However, the exact nature of the progenitors of type Ia supernovae remains unclear, hindering our understanding of the Universe. Examples of detailed binary population synthesis studies are \cite{Li1997}, \cite{Han1995b}, \cite{Yungelson2000}, \cite{Han2004}, \cite{Wang2009b}, \cite{Meng2009} and \cite{Toonen2012}. These studies assume that a carbon-oxygen white dwarf can grow in mass via accretion from a main sequence companion star, a giant companion star, or a helium companion star, and reach the Chandrasekhar mass limit to explode as a type Ia supernova (SNe Ia)  (\citealt{Hoyle1960,Whelan1973}). The merger of two carbon-oxygen white dwarfs may also lead to a SN Ia explosion if the total mass is over the Chandrasekhar mass limit (\citealt{Iben1984,Webbink1984,Han1998}).  
The studies shed light on the properties of SNe Ia, e.g. their formation channels, their birth rates, their properties in different environments and their evolution with redshift.

X-ray binaries are binary stars luminous in X-rays.  The X-rays are produced from the energy released during accretion of matter from one component onto the other component, a neutron (NS) star or a black hole (BH). X-ray binaries with NSs as the accretors may evolve to become millisecond pulsars. Such high energy binaries have been a subject of active research for many decades. Significant progress in understanding their nature and origins has been made with the help of BPS. \cite{Podsiadlowski2002}
calculated the relevant binary evolutionary sequences, \cite{Pfahl2003} was able to carry out a BPS investigation on the matter immediately afterwards. The study followed a population of intermediate- / low-mass X-ray binaries (I/LMXBs) from the incipeint stage to the current epoch and finally to the remnant state when they become binary millisecond pulsars (BMPs). Meaningful comparison of the theoretical predictions and the observations of LMXBs and BMPs were carried out. \cite{Liu2006} and \cite{Shao2015,Shao2020} have done detailed and systematic BPS studies on faint X-ray sources, ultraluminous X-ray sources, black-hole X-ray binaries. Those studies contribute significantly to our understanding of those binaries.

LIGO and VIRGO, the ground-based interferometric detectors for gravitational waves, have detected the gravitational wave signals of the merging of double BHs, double NSs, and BH+NS binaries. LISA and TianQin, the space-based interferometric detectors, are planned to be launched in the near future. The foreground of gravitational waves due to Galactic binary stars are presented via a BPS approach by \cite{Webbink1998}, \cite{Nelemans2001b} and \cite{Yu2010}. The merging pairs have been investigated by \cite{Belczynski2002b,Belczynski2016,Belczynski2020}, and the masses, the spins and the merging rates of the compact objects are given.
\cite{Li2019} studied the formation of low-mass WD pairs and many of them will be detectable by LISA (\citealt{Li2020}).

\cite{DeDonder2004} were the first to make Galactic chemical 
evolution simulations with the inclusion of
detailed binary evolutions via BPS. They showed that binary evolution mainly affects the Galactic evolution of carbon and iron. The majority of low- and intermediate-mass interacting binaries have avoided the late stages of AGB and therefore produce less carbon. Some binaries evolve to
SNe Ia, and SNe Ia produce the majority of iron. The SNe Ia rate, the moment at which the first SNe Ia start to form, and the moment at which their rates reach a maximum are all sensitive to the adopted progenitor models of SNe Ia. The merging of NS+NS binaries and NS+BH binaries produces $r$-process elements (e.g. gold), and the merging rates depend critically on the evolution of binary populations. \cite{Kobayashi2006} and \cite{Kobayashi2009} carried out detailed chemical evolution studies
with the latest stellar nucleosynthsis yields, the inclusion of hypernovae,
type II supernovae, and SNe Ia. The evolution depends on the supernovae 
progenitor models.

Pioneering work has been done by the Yunnan Group (\citealt{Zhang2004,Zhang2005,Han2007}) and the Auckland Group (\citealt{Eldridge2009,Eldridge2012,Stanway2016}) to include the effects of binary interactions in the synthetic spectra of stellar populations. Those studies have shown that binary interaction can
significantly change the spectral energy distribution.
For example, \cite{Han2007} showed that the far-UV excess of early type galaxies, the source of which had not been identified for many decades, can be well reproduced by accounting for radiation from hot subdwarfs resulting from binary interactions. \cite{Chen2015} conducted a binary population synthesis study on accreting white dwarfs and explained the {existence of soft X-ray band extended emissions} in galaxies.
The inclusion of binary interactions in evolutionary population synthesis
studies has already had a big impact on the derived ages, masses, and star formation rates in the studies of galaxies.

The re-ionizing photons of the Universe were conventionally assumed to be from single massive stars, and it is well-known that the number of photons falls short of the required amount by a factor of a few. However, since the binary fraction of massive stars is as high as 70\% (\citealt{Sana2012}), binary interaction may lead to the formation of hydrogen-envelope-stripped stars and massive blue stragglers. Both produce significant amounts of ionizing photons 10-200 Myr after starburst, and the relative importance of these photons are amplified as they escape more easily. de Mink's group 
(\citealt{Gotberg2020,Secunda2020}) showed that stellar population synthesis models taking into account binary stellar evolution provide a sound physical basis for cosmic re-ionization.

\section{The General Picture of Binary Population Synthesis}

\begin{figure}
\begin{center}
\includegraphics[width=\textheight,angle=90]{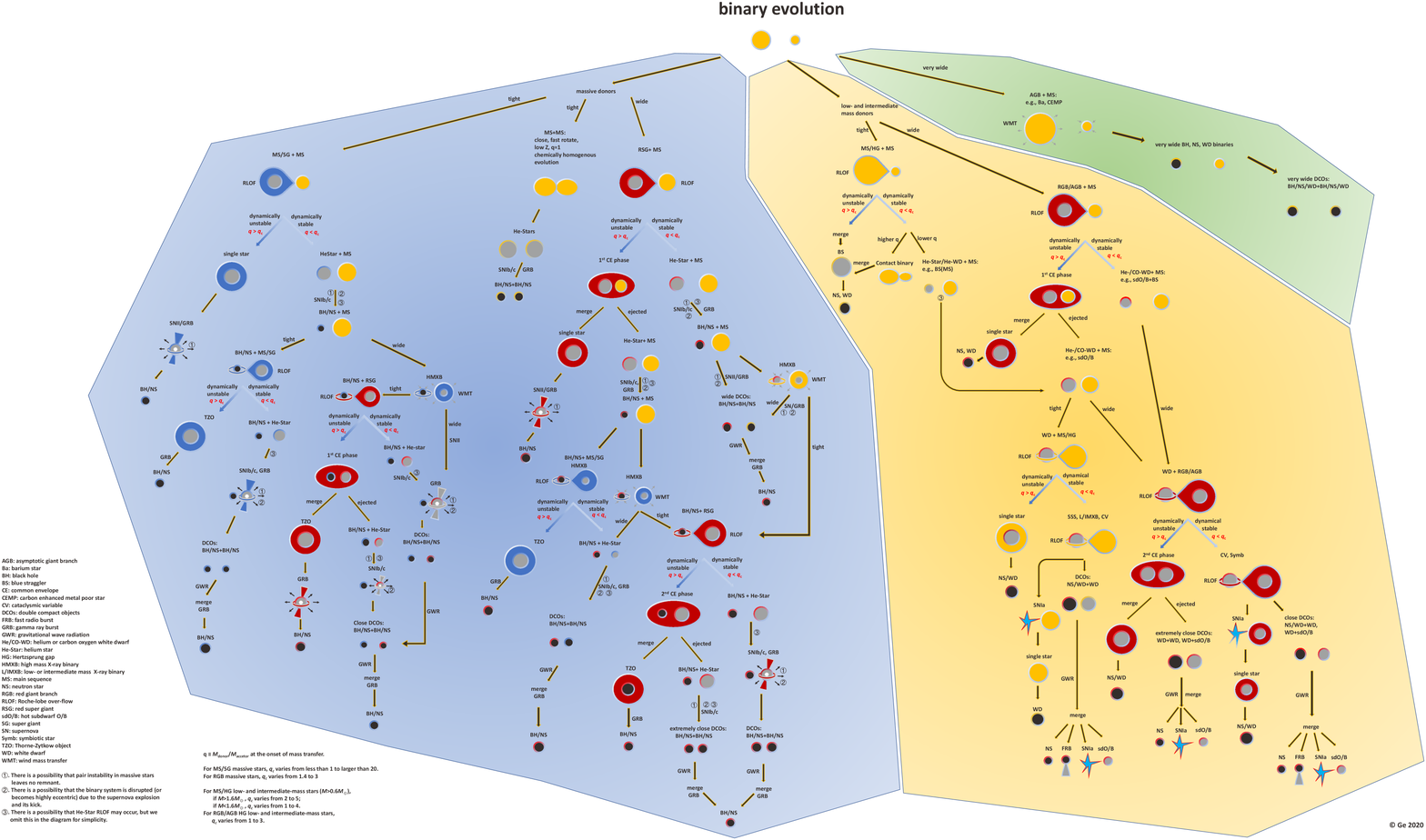}
\caption{Binary evolution tree.}
\end{center}
\label{flow}
\end{figure}

\begin{figure}
\begin{center}
\includegraphics[width=10cm,angle=0]{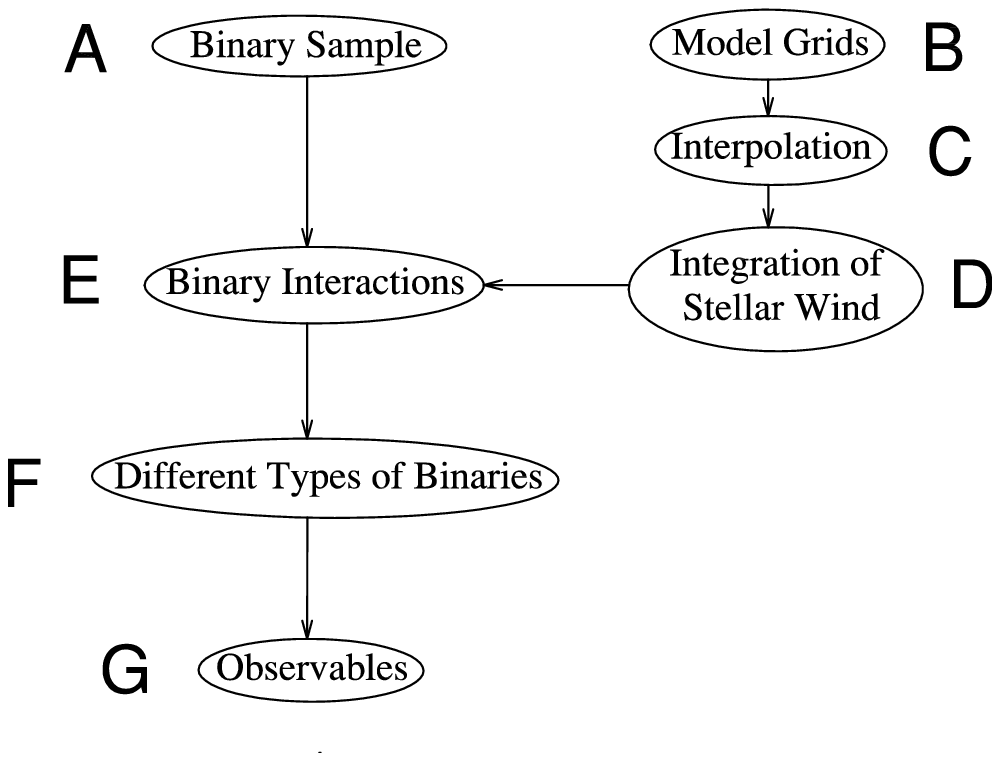}
\caption{Major steps in binary population synthesis studies.
Adopted from \cite{Han2003a} }
\end{center}
\label{code}
\end{figure}

Most exotic stars result from binary interactions.
Figure~1 is a binary evolution tree, showing how various exotic stars are formed. The tree can be much larger and much more complicated, and more channels need to be added for the formation of each particular kind of binary-related object not shown in the figure. 

Figure~2 illustrates the major steps in BPS studies. In a binary population synthesis study, we need to follow the evolution of a large number (say, a million) of binaries in a way similar to that of the evolution tree. Single stars are treated as binaries with wide orbits, and various binary interactions (e.g. those shown in Section 4.3)  are considered. Those interactions determine the fate of a binary and result in the formation of different types of objects. Both the properties of an individual exotic object and the statistical properties of those objects are compared to observations. From those comparisons, we can constrain the physical processes crucial for the formation of the previously mentioned exotic objects, elucidate their origin and explain their properties, and make predictions to be confirmed by future observations.

\section{Main Ingredients of Binary Population Synthesis}

\begin{figure}
\begin{center}
\includegraphics[width=\textwidth,angle=0]{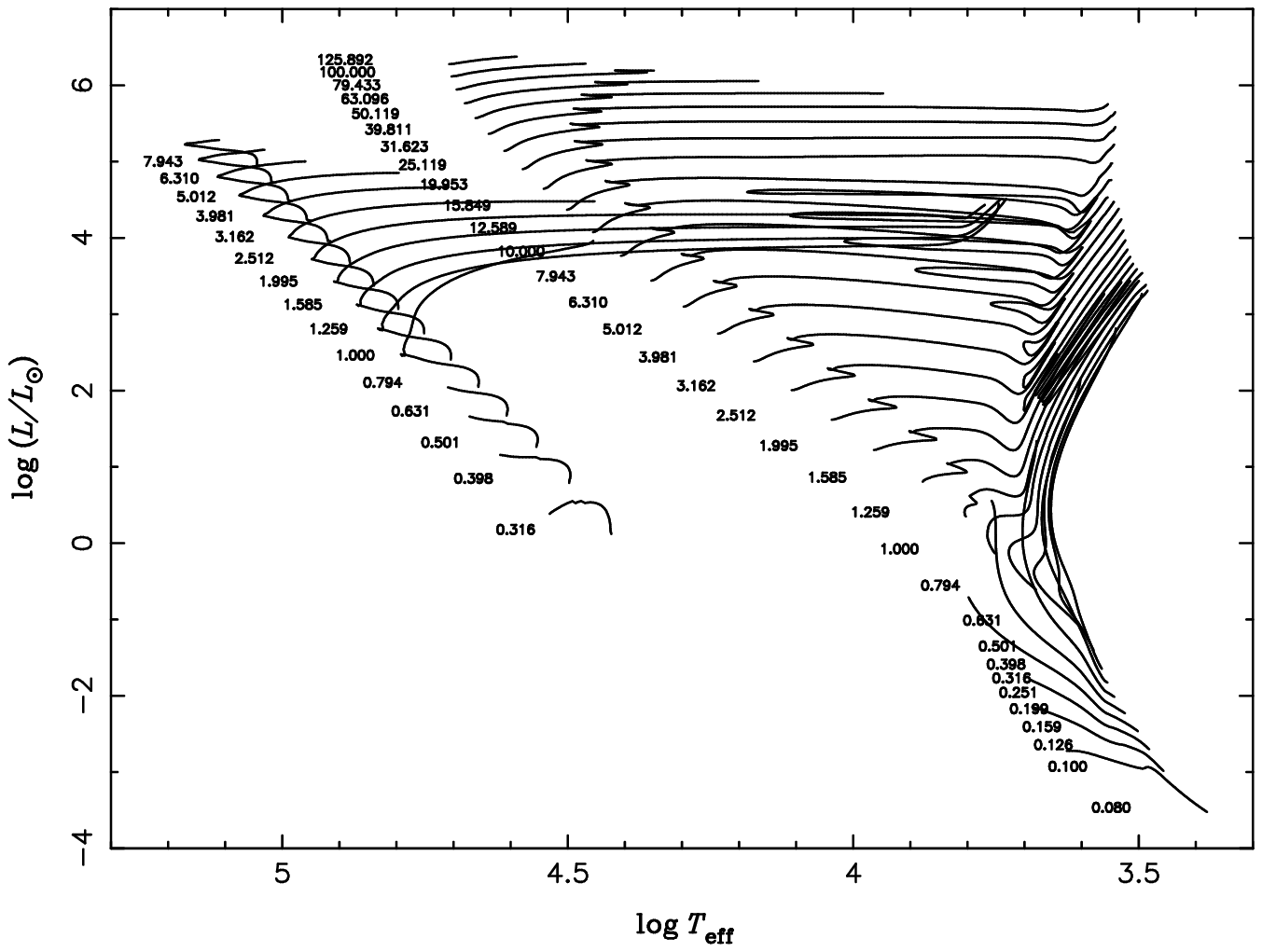}
\caption{Evolutionary tracks of Pop I stars. Masses in solar units are
given at the starting point of each track, regardless of whether the track is for the zero-age main-sequence (ZAMS) or the zero-age helium main-sequence (ZAHeMS). The helium stars are the sequence to the left.
Adopted from \cite{Han2001}}
\end{center}
\label{hrd}
\end{figure}

\subsection{Initial distributions of binaries}

In a BPS study, we need to generate sample binaries and then evolve them. To generate sample binaries, we need the star formation rate, the initial mass function of the primary, the mass ratio distribution, the orbital separation distribution, and the orbital eccentricity distribution. These are the basic inputs.

(1) We simply take a constant star formation rate over the last 15\, Gyr in most cases. 

(2) A \cite{Salpeter1955} initial mass function is usually adopted for analytic studies. In BPS studies, we usually adopt an initial mass function of \cite{Miller1979}, which is similar to that by \cite{Kroupa1993}. See \cite{Kroupa2001} for a canonical discussion. We generate the primary mass with the formula of \cite{Eggleton1989},
\begin{equation}
M_1={0.19X\over (1-X)^{0.75}+0.032(1-X)^{0.25}}, 
\end{equation}
where $X$ is a random number uniformly distributed between 0 and 1, and
the adopted ranges of primary masses are 0.8 to $126.0\,M_\odot$.

(3) We mainly take a constant mass-ratio distribution (for a discussion, see \citealt{Mazeh1992,Duchene2013}),
\begin{equation}
n(1/q)=1,\qquad  0\leq 1/q \leq 1, 
\end{equation}
where $q=M_1/M_2$. An alternative distribution of mass-ratio is the case where the masses of both binary components are chosen randomly and
independently from the same IMF.

(4) We simply assume that all stars are members of binary systems and that the distribution of separations is constant in $\log a$, where $a$ is the separation, for wide
binary separations, and falls off smoothly at close separations (for a discussion, see \citealt{Duquennoy1991,Duchene2013}),
\begin{equation}
an(a)=\cases {\alpha_{\rm sep}({a\over a_0})^m, & $a\leq a_0$;\cr
               \alpha_{\rm sep}, & $a_0 < a < a_1$\cr } 
\end{equation}
where $\alpha_{\rm sep} \approx 0.070$, $a_0=10\,R_\odot$,
$a_1=5.75\times 10^6\,R_\odot=0.13\,{\rm pc}$, and $m\approx 1.2$.
This distribution means that there is an equal number of wide binary
systems per logarithmic interval and that approximately 50\%
of stellar systems are binaries with orbital periods less than
100\,yr.

(5) We assume all binaries are circularized, i.e. $e=0$.

\subsection{Single Stellar Evolution}

To evolve a binary system, we need to follow the evolution of both components and deal with their interactions. The evolution of the components, i.e. the evolution of single stars, is calculated 
with fitted formulae, e.g. \cite{Tout1996} and \cite{Hurley2000},
or via interpolation in a stellar evolution model grid. 
Fitted formulae are easy to use, but stellar evolution model grids contain
more information and can be updated more conveniently.

Figure~3 shows two sets of stellar evolution models calculated with Eggleton's stellar evolution code (\citealt{Eggleton1967, Eggleton1971, Eggleton1972, Eggleton1973a, Eggleton1973b, Han1994, Pols1995}). The first set is
for a typical Pop I composition with hydrogen abundance $X=0.70$, helium 
abundance $Y=0.28$ and metallicity $Z=0.02$. These models do not include
mass loss, which can be parametrically dealt with afterwards. The models cover the range from $0.08M_\odot$ to $126.0M_\odot$ at roughly equal intervals of $0.1$ in $\log M$. The evolutionary tracks are terminated by the HD limit (\citealt{Humphreys1979,Lamers1988}) for 
massive stars, or when the stellar envelope has a positive binding energy for 
intermediate- or low-mass stars  (\citealt{Han1994,Meng2008}). 
The tracks for intermediate- 
or low-mass stars can also be terminated alternatively by the observed initial - final mass relations, which have quite a large scatter (\citealt{Weidemann2000,Ferrario2005}).
The second set of the tracks is for 
Pop I helium stars, with masses between $0.32M_\odot$ and $8.0M_\odot$.

\subsection{Binary Stellar Evolution}

\begin{figure}
\begin{center}
\includegraphics[width=10cm,angle=0]{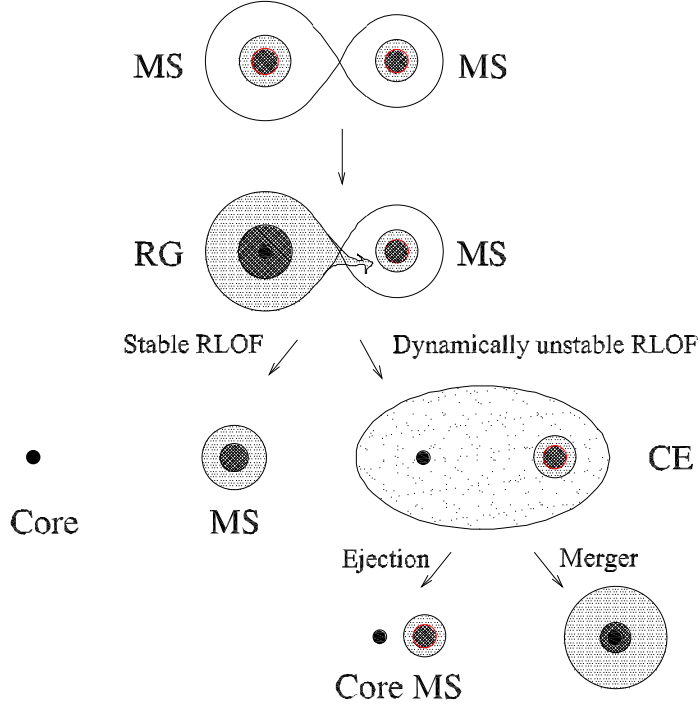}
\caption{Major binary interactions. Adopted from \cite{Han2003a}}
\end{center}
\label{binary}
\end{figure}

Binary stars interact in many ways, such as Roche lobe overflow (RLOF),
orbital angular momentum loss via gravitational wave radiation, magnetic braking, and stellar wind accretion. RLOF is the most important interaction, which changes the destiny of a binary system. A cartoon for major binary interactions is shown in Figure~4.

For a zero-age main sequence (ZAMS) binary star, the primary at first evolves and expands rather like a single star, until it fills its Roche lobe and starts to transfer its envelope mass to the secondary (see Figure~4). 
The Roche lobe radius $R_{\rm L}$ of the primary is given by \cite{Eggleton1983},
\begin{equation}
{R_{\rm L}\over A}= {0.49q^{2/3}\over 0.6q^{2/3} + \ln (1+q^{1/3})},
\qquad 0<q< \infty 
\end{equation}
where $A$ is the binary separation, $q={M_1\over M_2}$, 
and $M_1$ and $M_2$ are the masses of the primary and the
secondary, respectively. This approximation formula is accurate to $\sim 1\%$ and its derivative is smooth. Note that a similar approximating formula 
is given by \cite{Ge2020b} (their equations 33 and 34) for the volume-equivalent radii of the outer Lagrangian equipotential surface of the primary, the volume of which is that enclosed within the euqipotential surface passing through $L_2$ or $L_3$ and a plane passing through $L_1$ and perpendicular to the semi-major axis.

The mass transfer can be dynamically unstable, depending on the mass ratio of the primary to the secondary, on  the structure of the primary's envelope, and on the angular momentum 
loss of a binary system. Dynamical mass transfer will lead to the formation of a common envelope (CE, \citealt{Paczynski1976}). The CE engulfs the core of one star (primary) and its companion. The orbit of the embedded binary decays due to frictional drag and may release and deposit large amounts of energy into the envelope (\citealt{Livio1988}). The envelope may be ejected if the total deposited energy can overcome its binding energy.
For the CE ejection criterion, we introduced two model parameters, $\alpha_{\rm CE}$ for the common envelope ejection efficiency and $\alpha_{\rm th}$ for the thermal contribution to the binding energy of the envelope, which we write as
\begin{equation}
\alpha_{\rm CE}\,|\Delta E_{\rm orb}| > |E_{\rm gr} + \alpha_{\rm th}
\,E_{\rm th}|,
\end{equation}
where $\Delta E_{\rm orb}$ is the energy released from orbital contraction, $E_{\rm gr}$ the gravitational binding energy and $E_{\rm th}$ 
the thermal energy of the envelope.  
The orbital energy released is calculated with
\begin{equation}
\Delta E_{\rm orb}\simeq {GM_{\rm c}M_2\over 2a_{\rm f}}-{G(M_{\rm c}+
M_{\rm e})M_2\over 2a_{\rm i}},
\end{equation} 
where $M_{\rm c}$, $M_{\rm e}$, $M_2$ are the core mass and envelope mass of the primary, and the mass 
of the secondary, respectively, and $a_{\rm i}$ and $a_{\rm f}$ are the 
initial and final orbital separation, respectively. 
The binding energy $E_{\rm gr}$ and the thermal energy $E_{\rm th}$ of the envelope are obtained from full stellar structure calculations by 
\begin{equation}
E_{\rm gr}=\int_{M_{\rm c}}^{M_{\rm s}}{-{Gm\over r}} {\rm d}m
\end{equation} 
and 
\begin{equation}
E_{\rm th}=\int_{M_{\rm c}}^{M_{\rm s}}U {\rm d}m,
\end{equation} where $M_{\rm s}$ is the stellar surface mass and 
$M_{\rm c}$  the core mass. For the practical determination of 
$M_{\rm c}$ see Section 2 of \cite{Han1994}, and $U$ is the internal energy of the gas, involving terms due to the ionization of H and He and the dissociation of H$_2$, as well as the basic ${3\over 2}\Re T/\mu$ for a simple perfect gas, the energy of radiation, and the Fermi energy of a degenerate electron gas. 
CE ejection leads to the formation of a close binary with the core of the primary and its MS companion. The core may evolve further to form a WD star, a NS star or a BH.
If the CE fails to eject, the binary coalesces into a single fast-rotating star.

The mass transfer can be stable, i.e. the mass transfer is on a nuclear timescale or a thermal timescale. 
For a binary system with components $M_1$ and $M_2$ and separation $A$,
the orbital angular momentum $J_{\rm orb}$ is
\begin{equation}
J_{\rm orb}=\sqrt{GA\over M_1+M_2}M_1 M_2 \  
\end{equation}
Given that a mass $-{\rm d}M_1$ is lost from the primary, that a fraction 
$\beta$ of the lost mass is transferred to the
secondary star, that the remaining $-(1-\beta){\rm d}M_1$ is lost from the system, and that the mass lost from the system carries away the same specific angular momentum as pertains to the primary, the loss of angular momentum from the system 
is expressed as:
\begin{equation}
-{\rm d}J_{\rm orb} = {J_1\over M_1}(1-\beta){\rm d}M_1=
G^{1\over 2}M_2^2(M_1+M_2)^{-{3\over 2}}A^{1\over 2}(1-\beta){\rm d}M_1
\end{equation}
where $J_1$ is the angular momentum of the primary.
We then obtain the change of the separation due to mass transfer,
\begin{equation}
-{\rm d}\ln A=2{\rm d}\ln M_2 + 2 \beta {\rm d}\ln M_1 +
{\rm d}\ln (M_1+M_2) 
\end{equation}
Similarly, we obtain the change of the separation if we assume that the lost mass 
carries away the same specific angular momentum as pertains to the secondary,
\begin{equation}
-{\rm d}\ln A=2{\rm d}\ln M_1 + {2 \over \beta} {\rm d}\ln M_2 +
{\rm d}\ln (M_1+M_2) 
\end{equation}
We can also calculate the change if the loss mass carries away the angular momentum of 
the binary system,
\begin{equation}
-{\rm d}\ln A=2 \beta {\rm d}\ln M_1 + {2\over \beta} {\rm d}\ln M_2 -  {\rm d}\ln (M_1+M_2) 
\end{equation}

During binary evolution, some of the mass lost in the form of stellar wind from the primary may be accreted by the secondary star. The mass accretion rate $\dot M_2$ 
is expressed by \cite{Boffin1988} as,
\begin{equation}
\dot M_2=-{1\over \sqrt{1-e^2}} \left({GM_2\over 
V_{\rm wind}^2}\right)^2 {\alpha_{\rm acc} \dot M_1 
\over 2A^2 \left(1+{{V_{\rm orb}^2 \over V_{\rm wind}^2}}
\right)^{3/2}}  
\end{equation}
where $V_{\rm orb}=\sqrt{{G(M_1+M_2)\over A}}$ is the orbital velocity,
$V_{\rm wind}$ is the wind velocity at the position of the secondary star,
 $G$ is
the gravitational constant, $M_1$, $M_2$ and $A$ are the primary mass, secondary 
mass and orbital separation, respectively, $e$ is the orbital 
eccentricity and $\alpha_{\rm acc} $ is the accreting efficiency parameter.
We take
$\alpha_{\rm acc} = 1.5 $ 
(\citealt{Boffin1993}), which is appropriate for 
Bondi-Hoyle accretion.

The binary orbit decays due to gravitation radiation or magnetic braking.
The time scale $\tau_{\rm GR}$ (in yr)
for orbital angular momentum loss due to gravitation  radiation
(\citealt{Landau1962}) is expressed as,
\begin{equation}
\tau_{\rm GR} = {J_{\rm orb}\over \dot J_{\rm GR}} =
1.24\times 10^9 ({M_1\over M_\odot})^{-1}({M_2 \over M_\odot})^{-1}({M_1+M_2 \over M_\odot}
)^{-1}({A \over R_\odot})^4  
\end{equation}
where $\dot J_{\rm GR}$ is the orbital angular momentum loss rate due to 
gravitational radiation.

Magnetic braking results from magnetically coupled wind from the secondary star, and the braking exists if $M_2 > 0.37 M_\odot$.
The time-scale $\tau _{\rm MB}$ (in yrs)
 for orbital angular momentum loss due to magnetic braking 
(\citealt{Verbunt1981}) is expressed in a rather simple formula 
(\citealt{deKool1992}) as,
\begin{equation}
\tau _{\rm MB} = {J_{\rm orb} \over \dot J_{\rm MB}} =
4.5\times 10^6 ({M_1\over M_\odot}) ({M_1+M_2 \over M_\odot})^{-2}
({R_2 \over R_\odot})^{-\gamma}({A\over R_\odot})^5 
\end{equation}
where $\dot J_{\rm MB}$ is the orbital angular momentum loss rate due to
magnetic braking; $\gamma$ is usually taken to be 2.

In detailed binary evolution calculations, the angular momentum loss rate 
due to magnetic braking can be alternatively calculated
with a more complicated formula (\citealt{Sills2000}),

\begin{equation}
\dot{J}_{\rm MB}=\left\{\begin{array}{ll}
-K \omega^{3}\left(\frac{R_{2}}{R_{\odot}}\right)^{1 / 2}\left(\frac{M_{2}}{M_{\odot}}\right)^{-1 / 2}, & \omega \leq \omega_{\rm {crit }} \\
-K \omega_{\rm {crit }}^{2} \omega\left(\frac{R_{2}}{R_{\odot}}\right)^{1 / 2}\left(\frac{M_{2}}{M_{\odot}}\right)^{-1 / 2}, & \omega>\omega_{\rm {crit }}
\end{array}\right.
\end{equation}\\
where $K=2.7 \times 10^{47} \mathrm{g}\;\mathrm{cm}^{2}\;{\mathrm{s}}$, 
$\omega_{\rm crit}$ is the critical angular velocity, and
\begin{equation}
\omega_{\rm {crit }}(t)=\omega_{\rm {crit }, \odot} \frac{\tau_{t0, \odot}}{\tau_{t}}
\end{equation}
where $\omega_{\rm {crit}, \odot} = 2.9 \times 10^{-6} \mathrm{Hz}$, $\tau_{t0,\odot} \sim 28.4\;$days and $\tau_{t}$ are the global turnover timescales of the convective envelope for the Sun at its current age and that for the secondary star at age $t$, respectively.  They can be computed with the following equation.
\begin{equation}
    \tau_{t} = \int_{R_{\rm b}}^{R_{2}}\frac{{\rm d}r}{v_{\rm conv}}
\end{equation}
where $R_{\rm b}$ is the radius at the bottom of the surface convective envelope, and $v_{\rm conv}$ is the local convective velocity.

It is well known that neutron stars (NSs) receive a kick when they are born. One of the most compelling pieces of evidence for the existence of  NS kicks is the study by \cite{Lyne1994}, in which they found that the average velocity of young pulsars was $450\pm90\;$ km/s, much higher than that previously believed. One possible explanation for the NS kick is as follows. In the progenitors of core collapse supernovae, some hydrodynamical instabilities can result in asymmetric mass ejection, leading to the acceleration of the NS in the opposite direction of the ejecta (e.g. \citealt{Wongwathanarat2013,Janka2017}). With a deconvolution algorithm, \cite{Hobbs2005} inferred a velocity distribution from a sample of pulsars, which can be fitted by a Maxwell-Boltzmann distribution with $\sigma = 265\;$km/s (average speed of $\sim 420\;$km/s). 

Regarding the BH natal kick, it is poorly constrained and understood 
(e.g. \citealt{Willems2005, Fragos2009, Repetto2012,Repetto2015, Repetto2017}). A widely used model (e.g. \citealt{Belczynski2008}) for the BH natal kicks are that the NS-like kicks also work for BHs, but are scaled down linearly with the material-fallback fraction, i.e.
\begin{equation}
    V_{\rm kick} = V_{\rm kick, NS}(1-f_{\rm fb})
\end{equation}
Here, $V_{\rm kick, NS}$ is randomly chosen from the Maxwell-Boltzmann distribution with $\sigma = 265\;$km/s and $f_{\rm fb}$ is the material-fallback fraction. 

A binary may suffer a significant mass loss to produce a core collapse supernova, and the remnant may have undergone a kick at birth. If the mass loss or the kick velocity is very large, the binary orbit can become unbound and the two stars can consequently become single. Thermonuclear explosions, i.e. that of SNe Ia, leave a single star too in the case of the single degenerate scenario as its WD is completely destroyed. Some of those resultant single stars may be hypervelocity stars that escape from the Galaxy (\citealt{Wang2009a,Tauris2015,Neunteufel2020}). 
For a binary system with a circular orbit, its separation after explosion is given by
(see \citealt{Hills1983} and \citealt{Tauris2013} for details),
\begin{equation}
   \frac{A_{\rm f}}{A_{\rm i}} = \frac{1-(\Delta M/M_{\rm i})}{1-2(\Delta M/M_{\rm i})-(v_{\rm kick}/v_{\rm c})^{2}-2cos\theta(v_{\rm kick}/v_{\rm c})} 
\end{equation}
where $A_{\rm i}$ and $A_{\rm f}$ are the binary separation before and after the explosion, respectively. 
$\Delta M = M_{\rm i} -M_{\rm f}$ is the mass loss of the system, and $M_{\rm i}$ and $M_{\rm f}$ are the total masses of the binary system before and after the explosion. $v_{\rm kick}$ is the kick velocity of the remnant, $v_{\rm c} = \sqrt{GM_{\rm i}/A_{\rm i}}$ is the orbital velocity of the SN progenitor with respect to the companion star, and $\theta$ is the angle between the kick velocity and  pre-explosion orbital velocity vectors.  
The eccentricity after explosion is
\begin{equation}
e=\sqrt{1+\frac{2 E_{\mathrm{orb}} J_{\mathrm{orb}}^{2}}{\mu G^{2} M_{\mathrm{2f}}^{2} M_{\rm 1f}^{2}}}
\end{equation}
where $\mu$ is the reduced mass after explosion. The orbital energy of the system after explosion is given by
\begin{equation}
E_{\rm orb} = -\frac{GM_{\rm 1f}M_{\rm 2f}}{2A_{\rm f}},
\end{equation}
and the orbital angular momentum is
\begin{equation}
J_{\mathrm{orb}}=A_{\rm f} \mu \sqrt{\left(v_{\mathrm{c}}+v_{\rm kick} \cos \theta\right)^{2}+(v_{\rm kick} \sin \theta \sin \phi)^{2}}
\end{equation}
where $\phi$ is the angle between the pre-explosion orbital plane and the projection of the kick velocity vector onto a plane perpendicular to the pre-explosion velocity vector of the SN progenitor.

\section{Stability of Mass Transfer}

Mass transfer is the most important binary interaction, and its stability determines the fate of a binary system. The problem has been studied for many decades, and significant progress has been made recently. Below we discuss the stability criteria for mass transfer. 

The stability criteria for mass transfer rely on physical models or simulations of the mass transfer process. 
To obtain the criteria, we need to understand the response of the donor star to its mass loss (or mass transfer), the response of the orbit, and the response of the donor's Roche lobe. Whether the mass transfer is stable or not can be derived from the response of the donor's radius and its Roche-lobe radius to the mass loss. Three useful parameters, $\zeta_{\rm L}$, $\zeta_{\rm eq}$, and $\zeta_{\rm ad}$, are proposed by \cite{Webbink1985} to clarify how the binary system will evolve after the onset of the mass transfer. 

These three parameters, also called radius-mass exponents, describe the radius response of a donor star to mass loss:
\begin{equation}
\zeta_{\rm L}=\frac{{\rm \partial ln}R_{\rm L}}{{\rm \partial ln}M}\bigg|_{\rm RLOF},
\end{equation}
\begin{equation}
\zeta_{\rm eq}=\frac{{\rm \partial ln}R}{{\rm \partial ln}M}\bigg|_{\rm eq},
\end{equation}
\begin{equation}
\zeta_{\rm ad}=\frac{{\rm \partial ln}R}{{\rm \partial ln}M}\bigg|_{\rm ad},
\end{equation}
where $\zeta_{\rm L}$ is the response of the donor's Roche-lobe radius to mass loss, $\zeta_{\rm eq}$ is the response of the donor's radius to thermal-equilibrium mass loss, and $\zeta_{\rm ad}$ is the response of the donor's radius to adiabatic mass loss.
If $\zeta_{\rm ad} < \zeta_{\rm L}$, the donor star cannot retain hydrostatic equilibrium and mass loss proceeds on a dynamical timescale. If $\zeta_{\rm eq} < \zeta_{\rm L} < \zeta_{\rm ad}$, the donor star could retain hydrostatic equilibrium, but could not remain in thermal equilibrium and mass loss occurs on a thermal timescale. If $\zeta_{\rm L} < (\zeta_{\rm eq}, \zeta_{\rm ad})$, mass loss occurs due to evolutionary expansion of the donor star or due to the shrinkage of the Roche lobe owing to the angular momentum loss.

Useful insights into the behavior of donor stars undergoing adiabatic mass loss can be derived from simplified stellar models. \cite{Hjellming1987} investigated the properties of polytropic models with power-law equations of state for main-sequence stars and giant branch stars, and illuminate the qualitative behavior of those stars. \cite{Soberman1997} explored the stability criteria further for binary mass transfer, and investigated cases of non-conservative mass transfer. Aside from building a simplified physical model, one can also use classical stellar structure and evolution codes to simulate the binary mass transfer process, referred to as 
time-dependent mass loss models. For the time-dependent mass loss model, we need to make assumptions for the mass transfer rates, the mass loss, and the angular momentum loss from the binary system. The stability criterion is then derived by comparing the response of the donor radius and its Roche-lobe radius, or by examining whether the mass loss rate approaches a critical value. With stellar evolution codes, many authors derived the stability thresholds of stars on the Hertzsprung-gap (e.g. \citealt{Chen2003}), and on the giant branches 
(e.g. \citealt{Chen2008,Pavlovskii2015}).

The stability criteria for the mass transfer of binaries have been challenged by more and more observations.  For example, previous criteria implied that there exists a gap at around 1000 days for the orbital period distribution of post-AGB binaries. This is due to the fact that the critical mass ratio for stable mass transfer for binaries is less than one for giant branch star donors from polytropic models, and consequently the mass transfer leads to the formation of a common envelope, and the ejection of the common envelope results in the formation of binary systems with orbital periods much shorter than 1000 days. However, a large number of post-AGB binaries with orbital periods of around 1000 days have been observed. To solve the puzzle, we need to stabilize the above-mentioned mass transfer somehow.

To investigate the physics of mass transfer stability under such demands, \cite{Ge2010a,Ge2020a,Ge2020b} established a detailed stellar model for adiabatic mass loss, and a detailed stellar model for thermal equilibrium mass loss. Those models provide the asymptotic response of the donor star to different mass loss rates, i.e. those corresponding to the dynamical timescale mass transfer and to the thermal time scale mass transfer, respectively. 

With the adiabatic mass loss model, \cite{Ge2008,Ge2010a,Ge2010b,Ge2013,Ge2015,Ge2020a} 
carried out a systematic survey on the thresholds for dynamical timescale mass transfer over the entire span of possible evolutionary stages of donor stars. 
A detailed description is presented in the paper by \cite{Ge2020a}. We hereby summarize the results. 

\begin{figure}
 \begin{center}
 \includegraphics[width=\textwidth,angle=0]{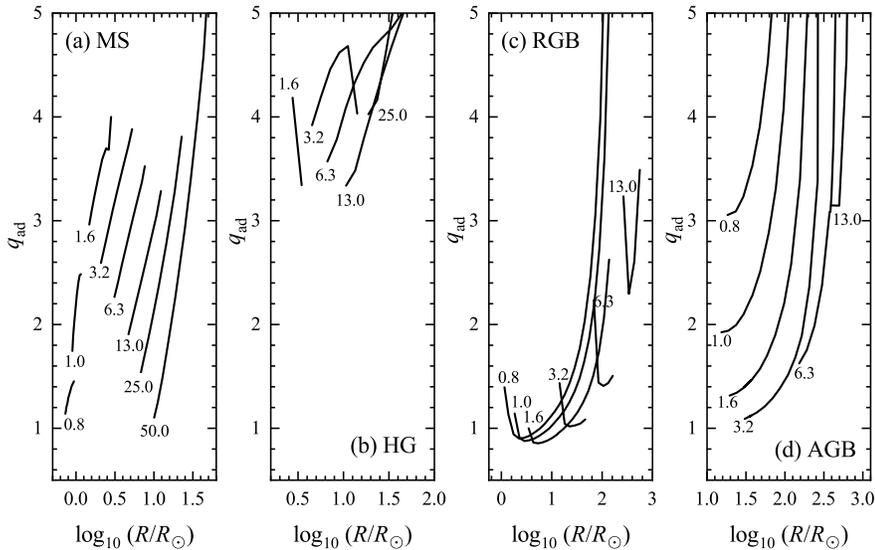}
 \caption{The critical mass ratio $q_{\rm ad}$ for dynamical timescale mass transfer as a function of radius of a donor star. The stellar radius increases as a star evolves, except for the turning points of the main sequence and the core-helium burning stages of low- and intermediate-mass stars. Panels (a), (b), (c), and (d) present the critical mass ratios of the donor stars on the main sequence, the Hertzsprung gap, the red-giant branch, and the asymptotic-giant branch, respectively. The masses of the donor stars are labeled at the starting points of each track (in solar units).}
 \label{qad}
\end{center}
\end{figure}

\subsection{Very Low-mass Stars}
For very low mass stars (with masses less than $\sim 0.5~M_\odot$), their convective envelopes dominate the response to rapid mass transfer. Furthermore, those  stars evolve very slowly, and remain almost unevolved on the main sequence within Hubble time. Therefore, the criteria from the complete and the composite polytropic models by \cite{Hjellming1987} present a good approximation. The critical mass ratio $q_{\rm ad}$ for dynamical timescale mass transfer of the very low-mass stars is approximated to 2/3.

\subsection{Main-sequence Stars}

Main-sequence stars with masses $\sim 1.6~M_\odot$ are in the transition zones between those which are dominated by convective envelopes and those with radiative envelopes. From panel (a) of Figure ~\ref{qad}, we obtain
the critical mass ratios as follows (see also \citealt{Ge2010a}). For main-sequence stars with masses less than $1.6~M_\odot$, the critical mass ratio $q_{\rm ad}$ decreases gradually with decreasing main-sequence mass to 2/3, which is for a completely convective $n=3/2$ polytrope. For main-sequence stars with masses larger than $1.6~M_\odot$, the critical mass ratio $q_{\rm ad}$ decreases gradually with increasing main-sequence mass. As the main-sequence stars with masses larger than $0.5~M_\odot$ evolve away from the zero-age main-sequence, the  masses of the radiative envelopes increases slightly for stars more masssive than 
$\sim 1.6M_{\odot}$ and that of the convective envelopes decrease slightly for stars less massive. Consequently, the critical mass ratio increases. Note that the critical mass ratio is not constant but changes dramatically as a star evolves on the main-sequence.

\subsection{Hertzsprung Gap Stars}

Stars with masses between  $\sim 1.6~M_\odot$ and $\sim 10~M_\odot$ evolve on a thermal timescale on the Hertzsprung gap, i.e. after their core-hydrogen is exhausted and before they reach the bottom of the red-giant branch or core helium burning. It is not the case for more massive stars, as the more massive stars start core helium burning already on the Hertzsprung gap. For stars with masses larger than $5.0~M_\odot$, the mass of their radiative envelopes increases monotonically on the Hertzsrpung gap. Therefore we see an evolution pattern of $q_{\rm ad}$ similar to that of main-sequence stars. For Hertzsprung gap stars, the critical mass ratio is $\sim 3$ and changes sharply as the stars evolve on the Hertzsprung gap, as shown in panel (b) of Figure~5. As discussed by \cite{Ge2015}, it is quite likely that the great majority of binaries with non-degenerate accretors and with mass ratios exceeding the value for delayed dynamical instability will in fact evolve into contact before the binaries actually reach the point of dynamical instability.

\subsection{Red-Giant Branch and Asymptotic Branch Stars}

The envelope changes from a radiation-dominated one to a convection-dominated one for stars at the bottom of the red-giant branch. 
Note also that thermal relaxation becomes increasingly important among luminous giants with extended envelopes. Panels (c) and (d) of Figure~5 show how critical mass ratios vary. The critical mass ratio 
$q_{\rm ad}$ decreases slightly and then starts to increase for stars with mass larger than $0.8~M_\odot$ on the red-giant branch. The critical ratio changes on the asymptotic-giant branch in a way similar to that on the red-giant branch, but the ratios are with higher values. For donor stars on the red-giant branch or on the asymptotic-giant branch with a deep enough convective envelope, the critical mass ratio becomes larger than $\sim 3$. This is because the envelopes of these stars are much more extended, the thermal timescale is very short (less than $10^2$ yr), and the dynamical timescale is even shorter. We also find that binaries with these luminous red giant donor stars may evolve into a common envelope phase through outer-Lagrangian overflow on a thermal timescale in the thermal equilibrium mass loss model (\citealt{Ge2020b}).

\section{Comparison with Observations}

With BPS,  we can obtain many properties of binary-related objects. The major properties include, but are not limited to, the birth rates, the numbers, the distributions of masses, the distribution of orbital periods, the distributions of mass ratios or mass functions, the distributions of spatial velocities, various statistical relations between the physical parameters, and the dependence of the properites on metallicity or redshift.
These properties depend somehow on the inputs and the assumptions of the relevant physical processes. Generally speaking, the dependence on the IMF and the initial orbital period distribution is weak. The dependence on the initial mass ratio distribution is modest, while the dependence on the CE evolution is very strong. It is obvious that the BPS results rely on the proposed formation scenarios of the binary-related objects and the relevant assumptions. By comparing the theoretical predictions of some observables of the objects with observations, we can constrain the model parameters and identify the formation scenarios for the relevant objects. A {\it best model} is then chosen accordingly. The best model is used to explain and reproduce other properties of the objects, and shed light on future observations and other related topics. 

\section{Future Prospects}

CE evolution is the key for the formation of many objects, but it remains the least understood process in binary evolution despite the many efforts made over the last few decades. \cite{Han1995b} first proposed that ionization energy may play a role in CE ejction, and three-dimensional hydrodynamical simulations taking into account ionization energy released (\citealt{Sand2020}) are making progress. Observations of post-CE 
binaries are in demand to constrain the CE process, especially those with well-defined progenitors. For example, observations of close hot subdwarf binaries
resulting from CE evolution at the tip of the first giant branch are viable probes of CE evolution 
(\citealt{Kupfer2015}). If we can observe such binaries in stellar clusters, the resulting constraint would be significant (\citealt{Han2008}).

It would be a good practice to study many kinds of binary-related objects simultaneously. We should not confine ourselves to studying a particular kind of objects. It is usually not too difficult to devise a scenario to repoduce observations of some specific type of stars, but it may be the case that the same assumptions, when applied to all stars, would lead to contradictions.

With the improvement of input physics and the advancement of computing power, BPS studies have evolved into a robust and precise science. Observational efforts provide a good estimate of binary fraction and precise statistical distributions of binary properties, and their dependence on metallicities (\citealt{Duchene2013,Moe2017,Moe2019,Liu2019}). 
Our understanding of stellar evolutionary processes is  being deepened, many physical processes are carefully treated,  and detailed evolution grids for single stars and even for binary stars are becoming available (see, for example, \citealt{Chen2014}). We will see a bigger impact of BPS studies on many aspects of astrophysics.     

\begin{acknowledgements}
This study is partly supported by the Natural Science Foundation of China (Nos 11521303, 11733008, 11673058, 11703081) and the Key Research Programme of Frontier Sciences of the CAS (No. ZDBS-LY-7005). We thank Dr. Matthias Kruckow for discussions in preparing the binary evolution tree (Figure~1).
\end{acknowledgements}

\label{lastpage}

\end{document}